\documentclass[12pt]{article}
\usepackage{subfigure}
\usepackage{latexsym}
\usepackage{amsmath}
\usepackage{amssymb}
\usepackage{amsthm}
\usepackage{undertilde}
\usepackage{graphicx}
\usepackage{hyperref}
\usepackage[latin1]{inputenc}
\usepackage[T1]{fontenc}
\usepackage[all]{xy}
\usepackage{lscape}
\makeatletter
\renewcommand{\@seccntformat}[1]
{\csname the#1\endcsname.\enspace} \makeatother
\def \XXint#1#2#3{{\setbox0=\hbox{$#1{#2#3}{\int}$}
\vcenter{\hbox{$#2#3$}}\kern-.5\wd0}}

\newtheorem{theorem}{Theorem}
\newtheorem{lemma}{Lemma}

\newtheorem{corollary}{Corollary}
\newtheorem{example}{Example}

\begin{document}
\begin{center}
{\bf Designing an Optimal Bonus--Malus System Using the Number of
Reported Claims, Steady-State Distribution, and Mixture Claim Size
Distribution}\\
{\sc Amir T. Payandeh Najafabadi\footnote{Corresponding author:
amirtpayandeh@sbu.ac.ir} \& Mansoureh Sakizadeh}\\
Department of Mathematical Sciences,
Shahid Beheshti University, G.C. Evin, 1983963113, Tehran, Iran.\\
\today
\end{center}
\begin{center}
\noindent {\bf Abstract}\\
\end{center}
This article, in a first step, considers two Bayes estimators for
the relativity premium of a given Bonus--Malus system. It then
develops a linear relativity premium that closes, in the sense of
weighted mean square error loss, to such Bayes estimators. In a
second step, it supposes that the claim size distribution for a
given Bonus--Malus system can be formulated as a finite mixture
distribution. It then evaluates the base premium under a Bayesian
framework for such a finite mixture distribution. The Loimaranta
efficiency of such a linear relativity premium, for several
Bonus--Malus systems, has been compared with two Bayes and
ordinary linear relativity premiums.
\section{Introduction}
A Bonus--Malus system is a popular actuarial tool that is based on
the true risk of policyholders and  categorizes them into a finite
number of levels, numbered from $1$ to $s.$ The main purpose of a
Bonus--Malus system is to determine the next year's premium of
classified policyholders based on their number of claims as well
as their current level. A Bonus--Malus system is a commercial and
practical version of a more general actuarial system, well known
as the rate-making system. The rate-making system determines the
next year's premium of each policyholder based on its average
number of claims in the last $t$ years. An ``optimal''
Bonus--Malus system can be designed through ({\bf 1})
pricing---i.e., evaluating both base and relativity premiums---of
a given Bonus--Malus system; ({\bf 2}) determining an
``appropriate'' statistical model for random variables involved in
a given Bonus--Malus system; ({\bf 3}) determining an ``optimal''
transition rule for a new Bonus--Malus system; and ({\bf 4}) a
combination of the above tasks. There is considerable attention
from researchers on the design of an optimal Bonus--Malus system
(or rate-making system). For instance, Lange (1969) provided some
useful actuarial and mathematical tools to determine the premium
of a given rate-making system. Dionne \& Vanasse (1992) employed
Poisson and negative binomial regression models to consider
available asymmetrical information whenever an actuary wants to
estimate the accident distribution in an insurance rate-making
system. Lemaire \& Zi (1994) compared 30 Bonus--Malus systems with
respect to 4 different criteria. Namely, they considered the
stationary average premium level, the coefficient of variation of
premiums, the efficiency of Bonus--Malus systems, and average
retention under the Bonus--Malus system as 4 appropriated measures
to study the optimality of a Bonus--Malus system. Based on these
comparison studies, they provided several practical suggestions to
design an appropriate Bonus--Malus system. Lemaire (1995) modelled
the claim frequency of a given Bonus--Malus system by a negative
binomial distribution and derived an estimate for premiums under a
quadratic loss function. Denuit (1997) employed a
Poisson--Goncharov distribution (introduced by Lefevre \& Picard,
1996) to model the annual number of reported claims under a given
Bonus--Malus system. Pinquet (1997) suggested the use of claim
severity in designing a Bonus--Malus system. Moreover, the
consideration of different types of claims has been suggested by
Pinquet (1998). Walhin \& Paris (1999) considered a finite mixture
Poisson distribution for random claim frequency of a given
Bonus--Malus system and derived a Bayesian premium. Denuit \&
Dhaene (2001) used an exponential loss function to calculate the
relativity premiums of a given Bonus--Malus system. Frangos \&
Vrontos (2001) considered both claim frequency and severity in
designing an optimal Bonus--Malus system. Morillo \& Berm\'udez
(2003) considered a Poisson--inverse Gaussian model to provide a
Bayesian relativity premium under an exponential loss function.
Boucher \& Denuit (2006) developed a Bayesian relativity premium
under a zero-inflated count model for panel data. In 2008, Boucher
\& Denuit extended Boucher \& Denuit's (2006) findings under
quadratic and exponential loss functions. Berm\'udez \& Morata
(2009) considered a Bonus--Malus system with two different types
of claims. They employed a bivariate Poisson regression model to
price such a Bonus--Malus system. In 2011, Berm\'udez \& Karlis,
based on work of Berm\'udez \& Morata (2009), developed a
situation in which the Bonus--Malus system has more than one type
of claim and there exists a non-ignorable correlation between such
types of claims. They used a Bayesian multivariate Poisson model
to price Bonus--Malus systems. Chen \& Li (2014) derived an
optimal linear relativity premium from the surplus of insurers'
viewpoints. Their linear relativity premium, somehow, can be
restated as a smoothing version of a Bayesian relativity premium
under a quadratic loss function. Payandeh Najafabadi et al. (2015)
employed the Payandeh Najafabadi (2010) method to derive a
credibility formula for the relativity premium of a given
rate-making system whenever count data have been sampled from a
zero-inflated Poisson gamma distribution. Teimourian et al. (2015)
employed the maximum entropy approach to determine a linear
relativity premium for a given Bonus--Malus system under both
long-run and short-run situations. In a Bonus--Malus system, each
level's premium is determined by multiplication of the base
premium and the corresponding relativity premium of that level.
Therefore, to determine the value of the premium in a Bonus--Malus
system, one must determine both base and relativity premiums. The
base premium has been evaluated using the size of the claim
regardless of the given Bonus--Malus system. To evaluate the
relativity premium, one must involve true but unobserved risk
characteristics (risk parameters) of the levels of the given
Bonus--Malus system. From a decision theory point of view, the
Bayes estimator offers an intellectual and acceptable estimation
for the relativity premium. Unfortunately, the Bayes estimator
suffers from the following disadvantages: ({\bf 1}) In most cases,
it cannot be restated as a convex combination of prior and current
observation means (see Payandeh Najafabadi 2010 for more details).
Therefore, for such cases, the Bayesian relativity premium, from a
computational viewpoint, is very time-consuming. ({\bf 2}) There
is no guarantee that the Bayesian relativity premium---say,
$r^{Bays}_j$ for $j=1,\cdots,s,$---satisfies the logical condition
$ a\leq r^{Bays}_1\leq r^{Bays}_2\leq \cdots\leq r^{Bays}_s\leq
b,$ where $a$ and $b$ are two given positive values. To eliminate
the above disadvantages, Gilde \& Sundt (1989), among others,
suggested the following linear class of estimators for the
relativity premium:
\begin{eqnarray}
\label{Linear-class-estimators}
\mathcal{C}:=\{r^{Lin}_l:=\alpha+\beta l,~\hbox{such that}~\alpha~ \&~ \beta\geq0~\hbox{and}~l=1,\cdots,s\}.
\end{eqnarray}
The two coefficients, $\alpha$ and $\beta$, have been determined by an
optimal criterion, such as minimizing the average square distance
between the relativity premium and the risk parameter.
The goals of this article are twofold. First, it is supposed that for
random claims, the given risk parameter $\eta$ under a Bonus--Malus
system can be reformulated as a finite mixture model. It then
provides an approximation for the Bayes estimator for the risk parameter
$\eta.$ This approximated Bayes estimator is very easy to
compute. Moreover, it does not suffer from the ``{\it
label-switching} problem''. Second, it considers two Bayes
estimators for the relativity premium, which are developed from the
distribution of {\it the number of reported claims} and {\it the
steady-state distribution} of the Bonus--Malus system.
Within the class of linear relativity premiums
\eqref{Linear-class-estimators}, it then develops a linear relativity
premium that simultaneously minimizes the square distances between
the linear estimator and these Bayes estimators. A practical
application of our finding, along with a comparison study, has
been given for some Bonus--Malus systems.
The rest of this article organized as follows. Section 2 collects
some preliminary result that play a vital role in this
article. The main results are represented in Section 3. Section 4
compares Loimaranta's efficiency for the optimal linear
relativity premium with the ordinary linear relativity premium and
Bayes relativity premiums for some given Bonus--Malus
systems. Concluding remarks along with some suggestions for future
research are given in Section 5.
\section{Preliminaries}
This section collects some primary results that will be used in the future.
For a complex-valued and integrable function $f,$ the Fourier
transform, say $\mathfrak{F}(f),$ and the inverse Fourier
transform, say $\mathfrak{F}^{-1}(f),$ are defined by
\begin{eqnarray*}
\mathfrak{F}(f;~x;~\omega) &=& \frac{1}{2\pi}\int_{\Re}f(x)e^{-ix\omega}dx \\
\mathfrak{F}^{-1}(f;~x;~\omega) &=& \int_{\Re}f(x)e^{ix\omega}dx,
\end{eqnarray*}
where $\omega\in{\Bbb R}.$ It is worth mentioning that the
well-known characteristic function for a random variable may be
viewed as the Fourier transform of the density/probability
function of this random variable. The \emph{Hausdorff--Young
theorem} states that an $L^q({\Bbb R})$ function $s$ and its
corresponding Fourier transform $\mathfrak{F}(s)\in L^{q^*}({\Bbb
R})$ satisfy $ ||\mathfrak{F}(s)||_{q^*}\leq (2\pi_*
)^{-1/q}||s||_q$, where $\pi_*$ stands for the Pi number and $1<
q\leq2$ and $1/q+1/{q^*}=1;$ see Pandey (1996) for more details.\\
Based upon standard distributions, the mixture models provide
statistical models that illustrate most aspects of complex
systems; see Tallis (1969) and McLachla \& Peel (2004), among
others, for more details on mixture models. Unfortunately, most
mixture models are not identifiable because they are invariant
under permutations of the indices of their components. This
identifiability problem is well known as the ``{\it
label-switching} problem''. The posterior distribution may also
inherit the ``{\it label-switching} problem'' from a prior
distribution that is also invariant under permutations (Rufo et
al., 2007). Under the ``{\it label-switching} problem'', there is
a positive probability that one of the components in the mixture
model does not contribute to any of the observations. Therefore,
the sample $x_1,\cdots,x_n$ has no information about this
component. Consequently, unknown parameter(s) of such a component
cannot be estimated under {\it either} classical {\it or} Bayesian
frameworks. A na{\"i}ve solution to the ``{\it label-switching}
problem'' is to impose some constraint on the parameter space for
the classical approach (Maroufy \& Marriott, 2015), and for the
Bayesian approach, some constraints have been added to the prior
distribution that lead to a posterior distribution that does not
suffer from the ``{\it label-switching} problem'' (Marin et al.,
2005). Unfortunately, insufficient care in the choice of suitable
identifiability constraints can lead to other problems (Rufo et
al., 2006). A random variable $X,$ given parameter $\eta,$ has a
finite mixture distribution with $k$ components if its
corresponding density function can be reformulated as
\begin{eqnarray}
\label{Mixture_distribution}
f_X(x|\eta) &=& \sum_{i=1}^{k}\upsilon_ig_i(x|\eta),
\end{eqnarray}
where $g_i(x|\eta)$-s are some given density functions,
$\upsilon_i\in[0,1],$ for $i=1,\cdots,k,$
$\sum_{i=1}^{k}\upsilon_i=1.$
Many authors have employed the mixture distribution in an actuarial
setting. For instance, Feldmann \& Whitt (1998) showed that a
large class of distributions, including several heavy tail
distributions, can be approximated by a finite mixture of exponential
distributions. Zhang \& Kwok (2010) approximated a given mixture
model with a simpler mixture model. They showed that this approach
increases computational time and, in several cases, improves the results
compared with other approximation methods. Bouguila (2011) employed a
finite mixture approximation method to model count data.
Payandeh Najafabadi (2015) approximated claim size distributions by
a finite mixture exponential distribution. He then provided an
accurate approximation for finite- and infinite-time ruin
probabilities for compound Poisson processes.
Suppose that there is a continuous random variable $X,$ where the given risk parameter
$\eta$ stands for the random claim size of a policyholder under an
insurance contract. Moreover, suppose that ({\bf 1}) the policyholder
under this insurance contract can be categorized into $s$ different
risk levels and ({\bf 2}) prior information about risk parameter
$\eta$ can be restated in terms of the following prior distribution
function: $\pi(\eta)=\sum_{i=1}^{s}\omega_i\pi_l(\eta).$
Therefore, the posterior distribution for $\eta$ given $X=x$ can
be restated as the following mixture posterior distribution.
\begin{eqnarray*}
\pi(\eta| X=x)&=&\frac{ f_{(X|\eta)}(x)\pi(\eta)}{\int_{0}^{\infty} f_{(X|\eta)}(x)\pi(\eta)d\mu(\eta)}=\frac{ f_{(X|\eta)}(x)\sum_{i=1}^{s}\omega_i \pi_i(\eta)}{\int_{0}^{\infty} f_{(X|\eta)}(x)\sum_{i=1}^{s}\omega_i \pi_i(\eta)d\mu(\eta)} \\ \\
&=&\sum_{i=1}^{s}\omega_i \frac{m_i(x)}{\sum_{i=1}^{s}\omega_i
m_i(x)}\pi_i(\eta| x) =\sum_{i=1}^{s}\rho_i \pi_i(\eta| x),
\end{eqnarray*}
where $m_i(x)=\int_{0}^{\infty}
\pi_i(\eta)f_{(X|\eta)}(x)d\mu(\eta)$ and
$\rho_i=\omega_im_i(x)/(\sum_{i=1}^{s}\omega_i m_i(x)).$
Consequently, the Bayes estimator for risk parameter $\eta$ under
the squared error loss function is
\begin{eqnarray*}
\delta_{Bayes}(x)&=& E(\eta| x)\\ &=&\int_{0}^{\infty} \eta
\sum_{i=1}^{s}\rho_i \pi_i(\eta| x) d\mu(\eta)\\
&=&\sum_{i=1}^{s}\rho_i \delta_{Bayes}^{\pi_i}(x).
\end{eqnarray*}
From the above result, it is concluded that the Bayes estimator
for risk parameter $\eta$ can be represented as a weighted
combination of the Bayes estimator for each category. The above
result cannot be generalized to a random sample size
$X_1,\cdots,X_n,$ $n(>1).$ In this situation, one must employ an
MCMC, a missing method, or a nonparametric Bayesian approach to
estimate the parameters of a mixture model under a Bayesian
framework. Unfortunately, all 3 of these approaches suffer from
the ``{\it label-switching} problem'' and are computationally very
time-consuming; see Marin et al. (2005) and Lin et al. (2014, \S
25), among others, for more details. Finding a closed form for the
likelihood function based on random sample $X_1,\cdots,X_n$ is the
main problem. The following provides a likelihood function (joint
distribution) for random sample $X_1\cdots,X_n,$ under a finite
mixture model.
\begin{theorem}
\label{minimal_Sufficient_Stat_mixture} Suppose random sample
$X_1\cdots,X_n,$ given risk parameter $\eta,$ is sampled
from the finite mixture density function
$f(t|\eta)=\sum_{i=1}^{k}\upsilon_ig_{i}(t|\eta),$ where $0\leq
\upsilon_i\leq 1$ and $\sum_{i=1}^{k}\upsilon_i=1.$ The
joint distribution function of random sample $X_1\cdots,X_n,$
given risk parameter $\eta,$ can then be restated as
\begin{eqnarray}
\label{Joint_density_mixture}
\prod_{j=1}^{n}f(x_j|\eta) &=&
\displaystyle\underbrace{\sum_{j_1=0}^{n}\cdots\sum_{j_k=0}^{n}}_{j_1+\cdots+j_k=n}\sum_{I_{j_1},\cdots,I_{j_k}}\upsilon_1^{j_1}g_1(I_{j_1}|\eta)\cdots
\upsilon_k^{j_k}g_k(I_{j_k}|\eta),
\end{eqnarray}
where $I_{j_1},\cdots,I_{j_k}$ (for $j_1,\cdots,j_k=0,\cdots,n,$
where $j_1+\cdots+j_k=n$) are distinct partitions of random
sample $X_1\cdots,X_n,$ with $j_1,\cdots, j_k$ elements,
respectively.
\end{theorem}
{\it Proof.} The desired result is obtained by partitioning random
sample $X_1\cdots,X_n$ into distinct partitions
$I_{j_1},\cdots,I_{j_k}.$ $\square$\\
Theorem
\eqref{minimal_Sufficient_Stat_mixture} provided an exact joint
density function of random sample $X_1\cdots,X_n,$ which are
sampled from a finite mixture distribution. Certainly, the above
finding cannot be employed in practical situations.

The following theorem studies a situation in which this joint
density function was approximated by a finite mixture
distribution. Hereafter, without loss of generality, we assume
that the weights of our finite mixture density functions are
equal. In a situation where some density functions have more
weight, such density functions can be repeated to achieve this
assumption.
\begin{theorem}
\label{Approximated_minimal_Sufficient_Stat_mixture} Suppose that
random sample $X_1\cdots,X_n,$ given risk parameter $\eta,$ is sampled from the finite mixture density function
$f(t|\eta)=\sum_{i=1}^{k}g_{i}(t|\eta)/k.$ Moreover, suppose that
the joint distribution function of random sample $X_1\cdots,X_n,$
given risk parameter $\eta,$ can be approximated by
\begin{eqnarray}
\label{Approximated_Joint_density_mixture}
\prod_{j=1}^{n}f(x_j|\eta) &\approx&
\displaystyle\frac{1}{k}\sum_{i=1}^{k}g_i(x_1,\cdots,x_n|\eta).
\end{eqnarray} The error bound for the above approximation then satisfies
\begin{eqnarray*}
\left|\prod_{j=1}^{n}f(x_j|\eta)-
\displaystyle\frac{1}{k}\sum_{i=1}^{k}g_i(x_1,\cdots,x_n|\eta)\right| &\leq &
\dfrac{M^k}{k^n}P(n),
\end{eqnarray*}
where $M=\max\{f_x,g_1,\cdots,g_k \}$ and $P(\cdot)$ stands for
the partition function.
\end{theorem}
\emph{Proof.} Using the result of Theorem
\eqref{minimal_Sufficient_Stat_mixture}, observe that
\begin{eqnarray*}
|\prod_{j=1}^{n}f(x_j|\eta)&-&
\displaystyle\frac{1}{k}\sum_{i=1}^{k}g_i(x_1,\cdots,x_n)|\\ &=& \left|\displaystyle\underbrace{\sum_{j_1=0}^{n}\cdots\sum_{j_k=0}^{n}}_{j_1+\cdots+j_k=n}\frac{1}{k^n}\sum_{I_{j_1},\cdots,I_{j_k}}g_1(I_{j_1}|\eta)\cdots
g_k(I_{j_k}|\eta)-\displaystyle\frac{1}{k}\sum_{i=1}^{k}g_i(x_1,\cdots,x_{n}|\eta)\right|\\
&\leq &\dfrac{M^k}{k^n} \left|\displaystyle\underbrace{\sum_{j_1=0}^{n-1}\cdots\sum_{j_k=0}^{n-1}}_{j_1+\cdots+j_k=n}\sum_{I_{j_1},\cdots,I_{j_k}} \right|= \dfrac{M^k}{k^n}
P(n).~ \square
\end{eqnarray*}
The following theorem studies a situation in which the density function of
a continuous random variable has been approximated by a finite
mixture density function.
\begin{theorem}
\label{Approximated_Joint_density_BY_mixture} Suppose that random
sample $X_1\cdots,X_n,$ given risk parameter $\eta,$ is
sampled from the density function $f(\cdot|\eta)$, and the density
function $f(\cdot|\eta)$ is approximated by the finite mixture
density function $\frac{1}{k}\sum_{i=1}^{k}g_i(\cdot|\eta).$
Moreover, suppose that the joint distribution function of random
sample $X_1\cdots,X_n,$ given risk parameter $\eta,$ is
approximated by
\begin{eqnarray*}
\prod_{j=1}^{n}f(x_j|\eta) &\approx& \displaystyle\frac{1}{k}\sum_{i=1}^{k}g_i(x_1,\cdots,x_{n}|\eta).
\end{eqnarray*}
An $L^p({\Bbb R})$-norm of the error bound for the above
approximation then satisfies
\begin{eqnarray*}
\left|\left|\prod_{j=1}^{n}f(x_j|\eta)-\prod_{j=1}^{n}\left(\frac{1}{k}\sum_{i=1}^{k}g_i(x_j|\eta)\right) \right|\right|_p &\leq
& \frac{nM^{n-1}}{k\sqrt[q]{2\pi_*}}\sum_{i=1}^{k}\left|\left|\psi_X(\cdot|\eta)-\psi_i(\cdot|\eta)\right|\right|_q,
\end{eqnarray*}
where $1<p\leq2,$ $1/q+1/{q}=1,$ $M=\max\{f_x,g_1,\cdots,g_k \},$ and $\psi_X(\cdot|\eta),$ $\psi_1(\cdot|\eta)\cdots\psi_k(\cdot|\eta)$ are the characteristic functions
corresponding to density functions $f_X(\cdot|\eta),$ $g_1(\cdot|\eta),\cdots,g_k(\cdot|\eta),$ respectively.
\end{theorem}
\emph{Proof.} For briefness, set
$f^*(\cdot|\eta):=\frac{1}{k}\sum_{i=1}^{k}g_i(\cdot|\eta).$
To obtain the desired result, employ the inequality
$|\prod_{j=1}^{n}f(x_j|\eta)-\prod_{j=1}^{n}f^*(x_j|\eta)|\leq
M^{n-1}\sum_{j=1}^{n}|f(x_j|\eta)-f^*(x_j|\eta)|$ (see Durrett,
2010, Lemma 3.4.3.) along with the triangle inequality, and observe
that
\begin{eqnarray*}
\left|\left|\prod_{j=1}^{n}f(x_j|\eta)-\prod_{j=1}^{n}f^*(x_j|\eta)\right|\right|_p &\leq
&
M^{n-1}\sum_{j=1}^{n}\left|\left|f(x_j|\eta)-f^*(x_j|\eta)\right|\right|_p.
\end{eqnarray*}
An application of the Hausdorff--Young theorem completes the
desired proof. $\square$

It is worth mentioning that an appropriate and practical
approximation for the density function $f(\cdot|\eta)$ by the
finite mixture density function
$\frac{1}{k}\sum_{i=1}^{k}g_i(\cdot|\eta)$ arrives at a situation
in which each $g_i(\cdot|\eta),$ for $i=1,\cdot,k,$ has a
dimensional minimal sufficient statistic, say $T_i(n),$ for risk
parameter $\eta$ based on random sample $X_1\cdots,X_n.$
Suppose that $\mathcal{BMS}$ stands for an ordinary Bonus--Malus system in
which a given policyholder moves between its $s$ levels, numbered
from 1 to s, according to the number of last year's reported claims and the
transition probability matrix ${\bf A}.$ Moreover, suppose that
$N_t,$ given risk parameter $\theta,$ stands for a counting
process that represents the number of reported claims by a policyholder
at year $t.$
Assuming $N_t,$ the given risk parameter $\theta$ is independent of the
level of a policyholder. Denuit et al. (2007, \S4) showed that the
transition probability matrix ${\bf A}$ can be reformulated as
\begin{eqnarray*}
A(\theta)&=&\sum _{n=0}^\infty T(n)P(N_t=n|\theta),
\end{eqnarray*}
where $T(n)=[t_{ij}(n)],$ for $i,j=1,\cdots,s,$ stands for a
matrix that describes the transition rules of the Bonus--Malus
system as follows: $t_{ij}(k)=1$ if by $k$ claims in a year, a
policyholder goes from level $i$ to $j$, and $t_{ij}(k)=0$ otherwise;
see Denuit et al. (2007, \S4) for more details.
In the situation where $N_t,$ given risk parameter $\theta,$ is
dependent on the level of a policyholder, the above result may be
extended as follows.
\begin{corollary}
Suppose that the number of reported claims at year $t$ for a policyholder
whose true level in a Bonus--Malus system $\mathcal{BMS}$ is $i$
is distributed according to the counting process $N_t$ with the
parameter $\theta_i.$ Then, $i\times j$ element of the transition
rules of the Bonus--Malus system $A,$ say $a_{ij}(\theta_i),$ is
$\sum _{n=0}^\infty t_{ij}(n)P(N_t=n|\theta_i),$ for
$i,j=1,\cdots,s.$
\end{corollary}
Assume that $L_t$ represents the level of a policyholder in year $t.$
Because the Markovian condition is met by stochastic process
$L_t,$ one may consider $L_t$ as a Markov chain with transition
probability matrix $A(\theta).$ Several authors discussed the
appropriateness of the ordinary Markov chain to model a given
Bonus--Malus system. For instance, Korolkiewicz \& Elliot (2008)
and Payandeh Najafabadi \& Kanani Dizaji (2011) employed a hidden
Markov model to study the behaviour of a given Bonus--Malus system.
Hereafter, we consider $L_t$ to be a Markov chain with the
transition probability matrix $A(\theta).$
The steady-state distribution for the Bonus--Malus system
$\mathcal{BMS}$ is presented as the long-run probabilistic
behaviour of $\mathcal{BMS}.$ The steady-state distribution is
a left-hand eigenvector of probability matrix ${\bf A}(\theta)$
with eigenvalue $1,$ ${\bf \pi}^{ss}
(\theta)=(\pi_1^{ss}(\theta),\cdots,\pi_s^{ss}(\theta))^\prime.$
Suppose that $L$ stands for the level occupied by a randomly selected
policyholder whenever the steady-state distribution is met
by the Bonus--Malus system $\mathcal{BMS}.$ Norberg (1976) showed that the
probability mass function for random variable $L$ can be restated
as
\begin{eqnarray}
\label{Dist.L} P(L=l)= \int_0^\infty \pi_l^{ss}(\theta)
dF_\Theta(\theta),
\end{eqnarray}
where $ \pi_l^{ss}(\cdot)$ and $F_\Theta(\cdot)$
stand for the steady-state distribution of level $l$ and the prior
distribution (structural function) for risk parameter $\theta$, respectively.

The relativity for a policyholder who occupied level $l,$ denoted
by $r_l,$ represents the amount of the base premium to be paid by
this policyholder. Certainly, the relativity premium for low-risk
policyholders is less than 1 (they received a bonus from the insurance
company), and it is greater than 1 for high-risk policyholders who
received a malus from the company. However, the relativity
premium $r_l$ must be satisfied: $a\leq r_1\leq r_2\leq\cdots\leq
r_s\leq b,$ where $a$ and $b$ are two given constants
determined by the insurance company to control the lowest and highest
premiums under the Bonus--Malus system.
The use of linear estimators and the use of Bayesian estimators are two well-known approaches to
estimating the relativity premium. The linear estimator is the
estimator within the class of \eqref{Linear-class-estimators} whose
coefficients have been estimated under optimal criteria.
Under the mean squared error optimality criteria, Gilde \& Sundt
(1999) showed that these coefficients are ${\hat
\alpha}^{Ord.Lin}=E(\Theta)-Cov(\Theta , L)E(L)/Var(L)$ and ${\hat
\beta}^{Ord.Lin}=Cov(\Theta , L)/Var(L).$
The Bayes estimator, under the squared error loss function for the
relativity premium, is obtained by minimizing the expectation of the
squared distance between the true relativity premium $\Theta$ and
its estimator $r_l.$ Such minimization can be achieved by
conditioning on {\it either} a random level of $L$ {\it or} a random
number of reported claims $N.$ The following provides such Bayes
estimators.
\begin{lemma}
\label{Bayes-Estimators_Relativity} Suppose that $\mathcal{BMS}$ stands
for an $s$-level Bonus--Malus system with the transition probability
matrix $A(\theta).$ Moreover, suppose the following:
\begin{description}
\item[1)] information on the number of reported claims $N_l,$ given $\theta_l,$ is
available, and the true relativity premium for a policyholder at level
$l$ is $\Theta_l.$ The Bayes estimator with respect to the prior
distribution $F_{\Theta_l}$ and under the squared error loss
function is then\begin{eqnarray}
r_l^{(1)} &:=& E(\Theta_l \vert N_l =n)= \frac{\int_{0}^{\infty} \theta P(N_l=n \vert \Theta_l=\theta\lambda)dF_{\Theta_l}(\theta) }{\int_{0}^{\infty} P(N_l=n \vert \Theta_l= \theta\lambda)dF_{\Theta_l}(\theta)
}.
\end{eqnarray}
\item[2)] $\Theta_l$ stands for the true relativity premium for a
randomly selected policyholder in level $l=1,\cdots,s.$ The Bayes estimator with respect to the prior
distribution $F_{\Theta_l}$ and under the squared error loss
function is then \begin{eqnarray}
r_l^{(2)} &:=& E(\Theta_l \vert L =l)= \frac{ \int_{0}^{\infty} \theta \pi_l^{ss}(\theta\lambda) dF_{\Theta_l}(\theta) }{ \int_{0}^{\infty} \pi_l^{ss}(\theta\lambda) dF_{\Theta_l}(\theta)
}
\end{eqnarray} whenever information on random level $L$ is considered,
and $\lambda$ stands for the a priori expected claim frequency.
\end{description}
\end{lemma}
{\it Proof.} Part (1) The desired results are obtained by conditioning
$E((\Theta_L-r_L)^2)$ on the random variable $N_l.$ For part (2), one must find the Bayes estimator by
minimizing $E((\Theta_L-r_L)^2).$ This estimator is obtained by
conditioning on the random variable $L.$ $\square$

In the situation where ({\bf 1}) $ N_l,$ given risk parameter
$\theta_l$, is distributed according to \emph{either} a
Poisson distribution \emph{or} a zero-inflated Poisson
distribution and ({\bf 2}) information about risk parameter
$\theta_l$ can be reformulated as $Gamma(a_l,b_l)$, the above
Bayes estimator $r_l^{(1)}$ can be simplified as $r_l^{(1)}=
(n+a_l)/(\lambda + b_l)$ for a Poisson distribution and
$r_l^{(1)}=[pb_l^{-a_l-1}+(1-p)(\lambda +
b_l)^{-a_l-1}]/[pb_l^{-a_l}+(1-p)(\lambda +
b_l)^{-a_l}]1_{\{0\}}(n)+[n+a_l]/[\lambda +
b_l]1_{\{1,2,\cdots\}}(n)$ for a zero-inflated Poisson
distribution. Moreover, under these assumptions, the Bayes estimator
$r_l^{(2)}$ can be simplified as
\begin{eqnarray*}
r_l^{(2)} &=& \dfrac{\int \theta e^{-\lambda \theta} (\lambda
\theta)^n e^{-b_l \theta} \theta^{a_l-1} \pi_l^{ss}(\theta) d
\theta}{\int e^{-\lambda \theta} (\lambda \theta)^n e^{-b_l
\theta} \theta^{a_l-1} \pi_l^{ss}(\theta) d \theta }
\end{eqnarray*}
The Loimaranta efficiency is a statistical tool that measures the
change of an expected premium paid by a policyholder subject to a
Bonus--Malus system as a function of its annual expected claim
frequency. The Loimaranta efficiency of an optimal Bonus--Malus
system increases with increasing annual expected claim frequency.
Greater respondence to increases in the annual expected
claim frequency represents greater appropriateness of the
Bonus--Malus system.
The Loimaranta efficiency $Eff_{Loi}$ for the annual expected claim
frequency $\vartheta$ is given by
\begin{eqnarray}
\label{Loimaranta_efficiency} Eff_{Loi}(\vartheta)=\frac{d\; ln
\bar{R}(\vartheta)}{d \; ln (\vartheta)},
\end{eqnarray}
where $\bar{R}(\vartheta)=\sum _{l=1} ^s r_l \pi_l^{ss}
(\vartheta);$ see Loimaranta (1972) for more details.
The Loimaranta efficiency measures how the average relativity
premium that must be paid by a policyholder who stays in a
Bonus--Malus system for a long time responds to the change of
annual expected claim frequency. An ideal efficiency should be
close to 1 for the most common values of annual expected claim
frequency $\vartheta.$ It is necessary to say that the Loimaranta
efficiency can be greater than 1; see De Pril (1978) for more
details.
\section{Main Results}
This section develops the base and relativity premiums for the given
Bonus--Malus system. Namely, the base premium has been evaluated
from a Bayesian framework, whereas the relativity premium is determined
through a linear approach. To develop a Bayes estimator for the
base premium, we suppose that the claim size random variable $X,$
given risk parameter $\eta,$ can be restated (approximately or
exactly) as a finite mixture distribution. Moreover, we suppose
that the prior information on risk parameter $\eta$ can be
reformulated as $s$ different prior distributions for $s$ classes of
the Bonus--Malus system. More precisely, the prior information on the risk
parameter $\eta$ can be restated as a mixture distribution
function with $s$ components.
\subsection{Bayesian approach to the base premium}
The Bonus--Malus system, based on the risk of policyholders,
categorized them into $s$ different risk classes. As mentioned
above, the premium of each class is determined by multiplying the
estimate of the risk parameter for the claim size by the estimate of the risk
parameter for the number of reported claims. In the ordinary approach
to evaluating the risk parameter for the claim size, say $\eta,$ the level of
the Bonus--Malus system is not considered, so we suppose that the random claim
size $X,$ given risk parameter $\eta,$ is distributed
according to a single (even unimodal) density function. Moreover,
we suppose that the prior information on risk parameter $\eta$ can be
reformulated as a single (even unimodal) prior distribution.
Certainly, policyholders' risk levels impact their claim size and
risk parameters. Therefore, these two assumptions will,
almost certainly, be violated in practice. To eliminate these two
barriers, this section supposes that both the claim size distribution and
the prior information of the risk parameter are two finite mixture
distributions. It then develops the Bayes estimator for risk
parameter $\eta.$
Unfortunately, for the joint distribution function of random sample
$X_1,\cdots,X_n,$ the given risk parameter $\eta$ cannot be restated in
closed form whenever the common density function is a finite
mixture distribution. Theorem
\eqref{Approximated_Joint_density_BY_mixture} provides an
approximation for this joint distribution function.

The Bayes estimator for the finite mixture model cannot be found
in a closed form, and one must employ an MCMC method, such a Gibbs
sampler (McLachlan \& Peel, 2004, \S4); a missing method; or a
nonparametric Bayesian approach to evaluate it numerically (Marin
et al., 2005 and Lin et al., 2014, \S 25). All three of these
approaches suffer from the ``{\it label-switching} problem'' and
are computationally very time-consuming; see Marin et al. (2005)
for more details.

The following provides an approximation for the Bayes estimator
under a finite mixture model. This approximated Bayes estimator is
very easy to compute and does not suffer from the ``{\it
label-switching} problem''.
\begin{theorem}
\label{Bayes_mixture_error_bound} Suppose that nonnegative random
sample $X_1\cdots,X_n,$ given risk parameter $\eta,$ is
sampled from the density function $f(\cdot|\eta).$ Moreover, suppose
that the joint density function $f_{X_1,\cdots,X_n}(\cdots|\eta)$
is approximated by the finite mixture density function
$f^*_{X_1,\cdots,X_n}(\cdots|\eta)=\frac{1}{k}\sum_{i=1}^{k}g_i(x_1,\cdots,x_{n}|\eta),$
where $T_i(n),$ for $i=1,\cdot,k,$ is a one-dimensional minimal
sufficient statistic for risk parameter $\eta$ based on random
sample $X_1\cdots,X_n$ with respect to the density function
$g_i(\cdot|\eta).$ Under the mixture prior distribution
$\pi(\eta)=\sum_{l=1}^{s}\omega_l\pi_l(\eta)$ and the
squared-error loss function, we have the following:
\begin{description}
\item[(1)] The Bayes estimator for $\eta$ can be approximated by
\begin{eqnarray}
\label{Bayes_mixture}
\nonumber \delta_{\pi;f}^{Bayes}(x_1\cdots,x_n) &\approx &
\delta_{\pi;f^*}^{Bayes}(x_1\cdots,x_n)\\ &=&
\sum_{i=1}^{k}\sum_{l=1}^{s}\rho_{i,l}(x_1,\cdots,x_{n})\delta_{\pi_l;g_i}^{Bayes}(T_i(n)),
\end{eqnarray}
where
$\rho_{i,l}(x_1,\cdots,x_{n})=\omega_lm_{i,l}(x_1,\cdots,x_{n})/(\sum_{i=1}^{k}\sum_{l=1}^{s}\omega_lm_{i,l}(x_1,\cdots,x_{n}))$
and
$m_{i,l}(x_1,\cdots,x_{n})=\int_{0}^{\infty}g_i(x_1,\cdots,x_{n})|\eta)\pi_l(\eta)d\eta$.
\item[(2)] The $L_p({\Bbb R})$-norm for the error bound of this approximation
satisfies
\begin{eqnarray}
\label{Err_Bound_Bayes_mixture}
\nonumber ||\delta_{\pi;f}^{Bayes}-\delta_{\pi;f^*}^{Bayes}||_p
&\leq &
\frac{nM^n}{km_p^2\sqrt[q]{2\pi_*}}\sum_{i=1}^{k}\sum_{l=1}\omega_l\int_0^\infty
\eta\pi_l(\eta)||\psi(\cdot|\eta)-\psi_i(\cdot|\eta)||_qd\eta\\
&& + \frac{nM^n
a}{km_p^2\sqrt[q]{2\pi_*}}\sum_{i=1}^{k}\sum_{l=1}\omega_l\int_0^\infty
\pi_l(\eta)||\psi(\cdot|\eta)-\psi_i(\cdot|\eta)||_qd\eta,
\end{eqnarray}
where $m_p=\min\{||\int_0^\infty f(\cdot|\eta)\pi(\eta)d\eta||_p, ||\int_0^\infty g_1(\cdot|\eta)\pi(\eta)d\eta||_p,\cdots, ||\int_0^\infty
g_k(\cdot|\eta)\pi(\eta)d\eta||_p\},$ $a=\int_0^\infty
\eta\pi(\eta)d\eta,$ $M=\max\{f,g_1,\cdots,g_k\},$
$\pi_*=3.141592654\cdots,$ $1/p+1/q=1,$ and $1\leq p\leq2.$
\end{description}
\end{theorem}
{\it Proof.} An application of Theorem
\eqref{Approximated_Joint_density_BY_mixture} completes the proof of
Part (i). Using Jensen's inequality (with an absolute-valued
function) along with the integral version of Minkowski's
inequality (Beckenbach \& Bellman, 2012, Page 22), one may
conclude that
\begin{eqnarray*}
||\delta_{\pi;f}^{Bayes}-\delta_{\pi;f^*}^{Bayes}||_p &\leq &
\left|\left|\int_{0}^{\infty}\eta\pi(\eta)\left|\dfrac{f_{X_1,\cdots,X_n}(\cdots|\eta)}{\int_{0}^{\infty}\pi(\eta f_{X_1,\cdots,X_n}(\cdots|\eta)d\eta} -\dfrac{f^*_{X_1,\cdots,X_n}(\cdots|\eta)}{\int_{0}^{\infty}\pi(\eta)f^*_{X_1,\cdots,X_n}(\cdots|\eta)d\eta}\right|d\eta
\right|\right|_p\\
&\leq & \int_{0}^{\infty}\eta\pi(\eta) \left|\left|\dfrac{f_{X_1,\cdots,X_n}(\cdots|\eta)}{\int_{0}^{\infty}\pi(\eta)f_{X_1,\cdots,X_n}(\cdots|\eta)d\eta} -\dfrac{f^*_{X_1,\cdots,X_n}(\cdots|\eta)}{\int_{0}^{\infty}\pi(\eta)f^*_{X_1,\cdots,X_n}(\cdots|\eta)d\eta}
\right|\right|_p d\eta.
\end{eqnarray*}
Using the extended Jensen's inequality for $L_p$-norm\footnote{The
ordinary Jensen's inequality states that for the convex function
$\phi(\cdot),$ one may conclude that
$\phi(||f||_1)\geq||\phi(f)||_1.$ Setting $g(\cdot)=|f(\cdot)|^p$
and $\phi(t)=\phi((t^p)^{1/p}),$ the ordinary Jensen's inequality
can be extended to $\phi(||f||_p)\geq||\phi(f)||_p$ for $1\leq
p\leq 2.$} (with $\phi(t)=1/t$ for t>0) as well as the
triangle inequality, the above inequality can be simplified as
\begin{eqnarray*}
||\delta_{\pi;f}^{Bayes}-\delta_{\pi;f^*}^{Bayes}||_p &\leq &
\frac{M}{m_p^2}\int_{0}^{\infty}\eta\pi(\eta)\left|\left|f_{X_1,\cdots,X_n}(\cdots|\eta)-f^*_{X_1,\cdots,X_n}(\cdots|\eta)\right|\right|_pd\eta\\
&&+\frac{Ma}{m_p^2}\int_{0}^{\infty}\pi(\eta)\left|\left|f_{X_1,\cdots,X_n}(\cdots|\eta)-f^*_{X_1,\cdots,X_n}(\cdots|\eta)\right|\right|_p
d\eta.
\end{eqnarray*}
The desired results will now be obtained by an application of
Theorem \eqref{Approximated_Joint_density_BY_mixture}. $\square$

To show the practical application of Theorem
\eqref{Bayes_mixture_error_bound}, two examples are now provided.
\begin{example}
\label{Example1_Bayes_mixture} Suppose that the random sample claim size
$X_1,\cdots,X_n,$ given risk parameter $\eta,$ is
distributed according to the following finite mixture
distribution.
\begin{eqnarray*}
f_X(x) &=& \frac{1}{3}LogNormal(\eta,1)+\frac{1}{3}LogNormal(\eta,1)+\frac{1}{3}Normal(\eta,1).
\end{eqnarray*}
Moreover, suppose that the prior information about risk parameter
$\eta$ can be reformulated as $\pi(\eta),$ with support
$[0,\infty).$
\end{example}
Using Theorem \eqref{Bayes_mixture_error_bound}, one may show that
the Bayes estimator for the risk parameter $\eta$ (and consequently the
base premium) is {\footnotesize
\begin{eqnarray*}
\delta_{\pi,f}^{Bayes}(x_1,\cdots,x_n) &=&\dfrac{2\int_{0}^{\infty}\eta\pi(\eta)\exp\{-\frac{1}{2}\left(T_1-2\eta
T_2+2T_2+n\eta^2 \right)
\}d\eta+\int_{0}^{\infty}\eta\pi(\eta)\exp\{-\frac{1}{2}\left(T_3-2\eta
T_4+n\eta^2 \right)
\}d\eta}{2\int_{0}^{\infty}\pi(\eta)\exp\{-\frac{1}{2}\left(T_1-2\eta
T_2+2T_2+n\eta^2 \right)
\}d\eta+\int_{0}^{\infty}\pi(\eta)\exp\{-\frac{1}{2}\left(T_3-2\eta
T_4+n\eta^2 \right) \}d\eta},
\end{eqnarray*}}\normalsize
where $T_1=\sum_{j=1}^{n}ln^2(x_j),$ $T_2=\sum_{j=1}^{n}ln(x_j),$
$T_3=\sum_{j=1}^{n}x^2_j,$ and $T_4=\sum_{j=1}^{n}x_j.$
\begin{example}
\label{Example2_Bayes_mixture} Suppose that the random sample claim size
$X_1,\cdots,X_n,$ given risk parameter $\eta,$ is
distributed according to the following the finite mixture
distribution.
\begin{eqnarray*}
f_X(x) &=& \frac{1}{2}Gamma(2,\eta)+\frac{1}{2}Pareto_{TypeI}(0.3,\eta).
\end{eqnarray*}
Moreover, suppose that the prior information about risk parameter
$\eta$ can be reformulated as $\pi(\eta),$ with support
$[0,\infty).$
\end{example}
Using Theorem \eqref{Bayes_mixture_error_bound}, one may show that
the Bayes estimator for the risk parameter $\eta$ (and consequently the
base premium) is {\footnotesize
\begin{eqnarray*}
\delta_{\pi,f^*}^{Bayes}(x_1,\cdots,x_n) &=&\dfrac{\int_{0}^{\infty}\eta^{2n+1}\pi(\eta)\exp\{T_2-\eta T_4\}d\eta+\int_{0}^{\infty}\eta^{n+1}(0.3)^{2\eta}\pi(\eta)\exp\{-(\eta+1)T_2\}I_{[0.3,\infty)}(x_{(1)})d\eta}{\int_{0}^{\infty}\eta^{2n}\pi(\eta)\exp\{T_2-\eta T_4\}d\eta+\int_{0}^{\infty}\eta^{n}(0.3)^{2\eta}\pi(\eta)\exp\{-(\eta+1)T_2\}I_{[0.3,\infty)}(x_{(1)})d\eta},
\end{eqnarray*}}\normalsize
where $T_2=\sum_{j=1}^{n}ln(x_j),$ $T_4=\sum_{j=1}^{n}x_j,$
$x_{(1)}=\min\{x_1,\cdots,x_n \},$ and $I_A(x)$ stands for the
indicator function.
\subsection{An optimal linear relativity premium}
From a decision theory point of view, the Bayes estimator offers
an intellectual and acceptable estimation for the relativity
premium. Unfortunately, two Bayes estimators for the relativity
premium, given by Lemma \eqref{Bayes-Estimators_Relativity}, are
computationally very time-consuming, and there is no guarantee
that such estimators satisfy logical condition $a\leq r_1\leq
r_2\leq \cdots\leq r_s\leq b;$ see Denuit et al. (2007) for more
details. To eliminate the above restrictions, Gilde \& Sundt
(1999) suggested the linear estimator $r_l^{Lin}$, within class
$\mathcal{C}$ given by \eqref{Linear-class-estimators}, for the
relativity premium, which is the minimized mean square error
$E(\Theta-r_L^{Lin})^2.$ The following theorem employs the
weighted mean square error method to provide the linear estimator
for the relativity premium, which is simultaneously close to both
Bayes estimators given by Lemma
\eqref{Bayes-Estimators_Relativity}.
\begin{theorem}
\label{Optimal_linear_Relativity} Suppose that $\mathcal{BMS}$ stands
for an $s$-level Bonus--Malus system with transition probability
matrix $A(\theta).$ Moreover suppose ({\bf 1}) that the number of
reported claims by a policyholder in level $l,$ say $N_l,$ with given
risk parameter $\theta_l,$ is distributed according to the
given probability mass function $P(N_l=n|\theta_l)$ and ({\bf 2})
prior information about the risk parameter $\theta_l$ is
reformulated by the cumulative distribution $F_{\Theta_l}.$
Within the class of linear estimator $\mathcal{C}$, the linear
relativity premium
\begin{eqnarray}
\label{Formula_Optimal_linear_Relativity}
r_l^{opt} &=& \alpha^{opt}+\beta^{opt}l
\end{eqnarray}
minimized the weighted mean square distance between Bayesian
relativity estimators $ r_l^{(1)}$ and $ r_l^{(2)},$ where
\begin{eqnarray*}
\alpha^{opt}&=&\xi E(r_L^{(1)})+(1-\xi)E(r_L^{(2)}) -\dfrac{E(L)}{Var(L)}\left[\xi Cov(L,r_L^{(1)})+(1-\xi)Cov(L,r_L^{(2)})\right] \\\\
\beta^{opt}&=&\dfrac{1}{Var(L)}\left[\xi Cov(L,r_L^{(1)})+(1-\xi)Cov(L,r_L^{(2)})\right]
\end{eqnarray*}
$E(r_L^{(1)}) =\sum_{l=1}^s r_l^{(1)}P(L=l),$
$E(r_L^{(2)})=\sum_{l=1}^s r_l^{(2)} P(L=l),$ $Cov(L,r_L^{(1)})=
\sum_{l=1}^s l \int_{0}^{\infty} r_l^{(1)}\pi_l^{ss}(\theta)
dF_{\Theta_l}(\theta) -E(r_L^{(1)}) E(L),$
$Cov(L,r_L^{(2)})=\sum_{l=1}^s l \int_{0}^{\infty}
r_l^{(2)}\pi_l^{ss}(\theta) dF_{\Theta_l}(\theta)-E(r_L^{(2)})
E(L),$ $\xi$ is a given number in $[0,1],$ and $ r_l^{(1)}$ and $
r_l^{(2)}$ are given by Lemma \eqref{Bayes-Estimators_Relativity}.
\end{theorem}
{\it Proof.} The weighted mean square distance between Bayesian
relativity estimators $ r_l^{(1)},$ $ r_l^{(2)}$ and linear
relativity premium $r_l^{Lin}$ within class $\mathcal{C},$ given
by \eqref{Linear-class-estimators}, can be restated as
\begin{eqnarray*}
WMSE(\alpha, \beta) &=& \xi
E(r_L^{(1)}-r_L^{Lin})^2+(1-\xi)E(r_L^{(2)}-r_L^{Lin})^2,
\end{eqnarray*}
where $\xi$ is a given number in $[0,1].$ Setting the derivative
of $WMSE(\alpha, \beta)$ with respect to $\alpha$ (and with
respect to $\beta$) equal to 0 yields $\alpha^{opt}$ and
$\beta^{opt}.$ To show that these $\alpha^{opt}$ and $\beta^{opt}$
minimize $WMSE(\alpha, \beta),$ one must show that its
corresponding Hessian matrix is positive semi-definite. This can be achieved by showing that the trace and determinate of the
Hessian matrix are nonnegative. Because $\partial^2 WMSE(\alpha,
\beta)/\partial\alpha^2=1,$ $\partial^2 WMSE(\alpha,
\beta)/\partial\alpha\partial\beta=\partial^2 WMSE(\alpha,
\beta)/\partial\beta\partial\alpha=E(L),$ $\partial^2 WMSE(\alpha,
\beta/\partial\beta^2=E(L^2),$ one may show that the trace and
determinate of the Hessian matrix are $1+E(L^2)\geq0$ and
$E(L^2)-E^2(L)=Var(L)\geq0,$ respectively. This observation completes the
desired results. $\square$
\section{Practical applications}
This section considers the Bonus--Malus system of Ireland (Hong
Kong), Kenya, and Brazil to show the application of our
findings.\footnote{Because these three Bonus--Malus systems have
been studied by Lemaire \& Zi (1994), we reconsider them for our
study. It is worth mentioning that our results can be
employed for any Bonus--Malus system in which policyholders move
between its levels according to their number of reported claims.}
Table 1 shows such Bonus--Malus systems.
\begin{center}
Table 1: Bonus--Malus systems for three countries.\\
\begin{tabular}{ c c c c}
\hline
Country & Number of classes & Starting level & Scale \\
\hline
Ireland (Hong Kong) & 6 & 6 & -1/ +3 \\
Kenya & 7 & 7 & -1 /Top\\
Brazil & 7 & 7 & -1/+1\\
\hline \\
\end{tabular}
\end{center}
Two base and relativity premiums for the Bonus--Malus systems given
in Table 1 have been evaluated using the methods developed above.
\subsection*{Relativity premium}
To evaluate the relativity premium for these Bonus--Malus systems,
we suppose that the number of reported claims for a policyholder at level
$l,$ say $N_l,$ given risk parameter $\theta_l,$ has been
distributed according to \emph{either} a Poisson distribution
\emph{or} a zero-inflated Poisson distribution. We then evaluate
the relativity premium using the ordinary linear approach (given
by Gilde \& Sundt, 1999), both Bayes estimators (given by
Lemma \ref{Bayes-Estimators_Relativity}) and the optimal linear
relativity premium (given by Theorem
\ref{Optimal_linear_Relativity}) whenever $\hat{\lambda}= 0.1474$
(Denuit et al., 2007, Page 91). These four estimators for different
values of $\lambda$ have been compared using the Loimaranta
efficiency, given by Equation \eqref{Loimaranta_efficiency}.
Tables 2 to 4 show four such estimators.

\begin{landscape}\begin{scriptsize}\begin{center}
Table 2: Relativity premium under Kenya's Bonus--Malus system.\\
\begin{tabular}{c c c c c c c c c c c c c}
\hline &&&& & $ N_l|\theta_l\sim Poisson(\theta_l)$ & & &&&$ N_l|\theta_l\sim ZIPoisson(\theta_l)$&&\\
\cline{5-8} \cline{9-13}
$l$ & $ \omega_l $ & $\pi_l$ & $ P(L=l) $ & $r_{l}^{(1)}$ & $r_{l}^{(2)}$ & $r_{l}^{Lin} $ & $r_{l}^{opt} $& $ P(L=l) $ &$r_{l}^{(1)}$ & $r_{l}^{(2)}$ & $r_{l}^{Lin} $ & $r_{l}^{opt} $\\
\hline
1 & $ \frac{1}{7} $ & Gamma(1,7) & 0.486 & 0.143 & 0.127 & 0.796 & 0.139 & 0.635 & 0.143 & 0.133 & 0.901 & 0.142\\
2 & $ \frac{1}{7} $ & Gamma(3,7) & 0.051 & 0.429 & 0.513 & 0.894 & 0.436 & 0.045 & 0.429 & 0.534 & 0.972 & 0.440\\
3 & $ \frac{1}{7} $ & Gamma(5,7) & 0.061 & 0.714 & 0.783 & 0.992 & 0.732 & 0.050 & 0.714 & 0.810 & 1.043 &0.737 \\
4 & $ \frac{1}{7} $ & Gamma(7,7) & 0.073 & 1.000 & 1.065 & 1.90 & 1.029 & 0.056 & 1.000 & 1.093 & 1.113 & 1.035\\
5 & $ \frac{1}{7} $ & Gamma(9,7) & 0.088 & 1.286 & 1.358 & 1.188 & 1.326 & 0.063 & 1.286 & 1.383 & 1.184 &1.333 \\
6 & $ \frac{1}{7} $ & Gamma(11,7) & 0.108 & 1.571 & 1.663 & 1.287 & 1.622 & 0.071 & 1.571 & 1.679 & 1.254 &1.631\\
7 & $ \frac{1}{7} $ & Gamma(13,7) & 0.133 & 1.857 & 1.980 & 1.385 & 1.919 & 0.080 & 1.857 & 1.980 & 1.325 & 1.928 \\
\hline
\end{tabular}\\
Table 3: Relativity premium under Hong Kong's Bonus--Malus system.\\
\begin{tabular}{c c c c c c c c c c c c c}
\hline &&&& & $ N_l|\theta_l\sim Poisson(\theta_l)$ & & &&&$ N_l|\theta_l\sim ZIPoisson(\theta_l)$&&\\
\cline{5-8} \cline{9-13}
$l$ & $ \omega_l $ & $\pi_l$ & $ P(L=l) $ & $r_{l}^{(1)}$ & $r_{l}^{(2)}$ & $r_{l}^{Lin} $ & $r_{l}^{opt} $& $ P(L=l) $ &$r_{l}^{(1)}$ & $r_{l}^{(2)}$ & $r_{l}^{Lin} $ & $r_{l}^{opt} $\\
\hline
1 & $ \frac{1}{6} $ & Gamma(1,6) & 0.699 & 0.167 & 0.158 & 0.165 & 0.917 &0.870 & 0.167 & 0.163 & 0.166 & 0.985\\
2 & $ \frac{1}{6} $ & Gamma(3,6) & 0.090 & 0.500 & 0.630 & 0.535 & 1.036 &0.045 & 0.500 & 0.645 & 0.50 & 1.041\\
3 & $ \frac{1}{6} $ & Gamma(5,6) & 0.109 & 0.833 & 0.961 & 0.904 & 1.154 &0.054 & 0.833 & 0.989 & 0.914 & 1.096\\
4 & $ \frac{1}{6} $ & Gamma(7,6) & 0.052 & 1.167 & 1.425 & 1.273 & 1.273 &0.016 & 1.167 & 1.460 & 1.288 & 1.152\\
5 & $ \frac{1}{6} $ & Gamma(9,6) & 0.017 & 1.500 & 1.701 & 1.643 & 1.392 &0.005 & 1.500 & 1.768 & 1.662 & 1.208\\
6 & $ \frac{1}{6} $ & Gamma(11,6) & 0.033 & 1.833 & 2.168 & 2.012 & 1.510 &0.009 & 1.833 & 2.156 & 2.036 & 1.263\\
\hline
\end{tabular}\\
Table 4: Relativity premium under Brazil's Bonus--Malus system.\\
\begin{tabular}{c c c c c c c c c c c c c}
\hline &&&& & $ N_l|\theta_l\sim Poisson(\theta_l)$ & & &&&$ N_l|\theta_l\sim ZIPoisson(\theta_l)$&&\\
\cline{5-8} \cline{9-13}
$l$ & $ \omega_l $ & $\pi_l$ & $ P(L=l) $ & $r_{l}^{(1)}$ & $r_{l}^{(2)}$ & $r_{l}^{Lin} $ & $r_{l}^{opt} $& $ P(L=l) $ &$r_{l}^{(1)}$ & $r_{l}^{(2)}$ & $r_{l}^{Lin} $ & $r_{l}^{opt} $\\
\hline
1 & $ \frac{1}{7} $ & Gamma(1,7) & 0.819 & 0.143 & 0.140 & 0.972 & 0.141 & 0.860 & 0.143 & 0.140 & 0.983 & 0.142 \\
2 & $ \frac{1}{7} $ & Gamma(3,7) & 0.117 & 0.429 & 0.561 & 1.068 & 0.499& 0.099 & 0.429 & 0.563 & 1.068 & 0.495 \\
3 & $ \frac{1}{7} $ & Gamma(5,7) & 0.038 & 0.714 & 0.981 & 1.163 & 0.846& 0.027 & 0.714 & 0.983 & 1.154 & 0.848\\
4 & $ \frac{1}{7} $ & Gamma(7,7) & 0.015 & 0.999 & 1.399 & 1.259 & 1.199 & 0.009 & 1.000 & 1.402 & 1.240 & 1.201\\
5 & $ \frac{1}{7} $ & Gamma(9,7) & 0.006 & 1.286 & 1.812 & 1.354 & 1.551 & 0.003 & 1.286 & 1.817 & 1.325 & 1.555\\
6 & $ \frac{1}{7} $ & Gamma(11,7) & 0.003 & 1.571 & 2.218 & 1.449 & 1.904 & 0.001 & 1.571 & 2.228 & 1.411 & 1.908\\
7 & $ \frac{1}{7} $ & Gamma(13,7) & 0.002 & 1.857 & 2.362 & 1.545 & 2.257 & 0.001 & 1.857 & 2.368 & 1.496 & 2.261\\
\hline
\end{tabular}
\end{center}\end{scriptsize}\end{landscape}

Figure 1 illustrates the behaviour of the Loimaranta efficiency for
four relativity premiums against the a priori expected claim
frequency $\lambda.$
\begin{center}
\begin{figure}[h!]
\centering\subfigure[]{
\includegraphics[width=9cm,height=6cm]{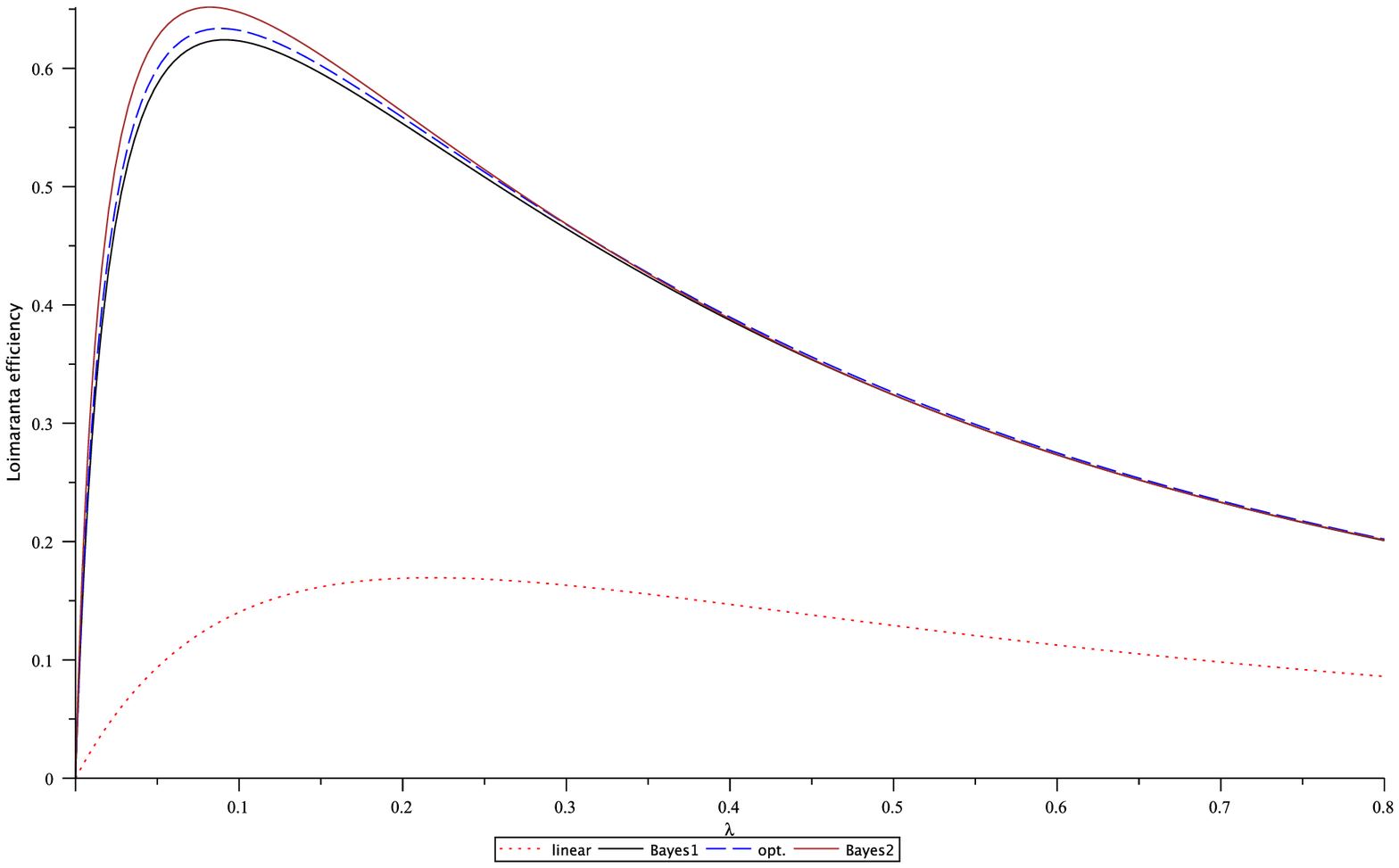}}\subfigure[]{
\includegraphics[width=9cm,height=6cm]{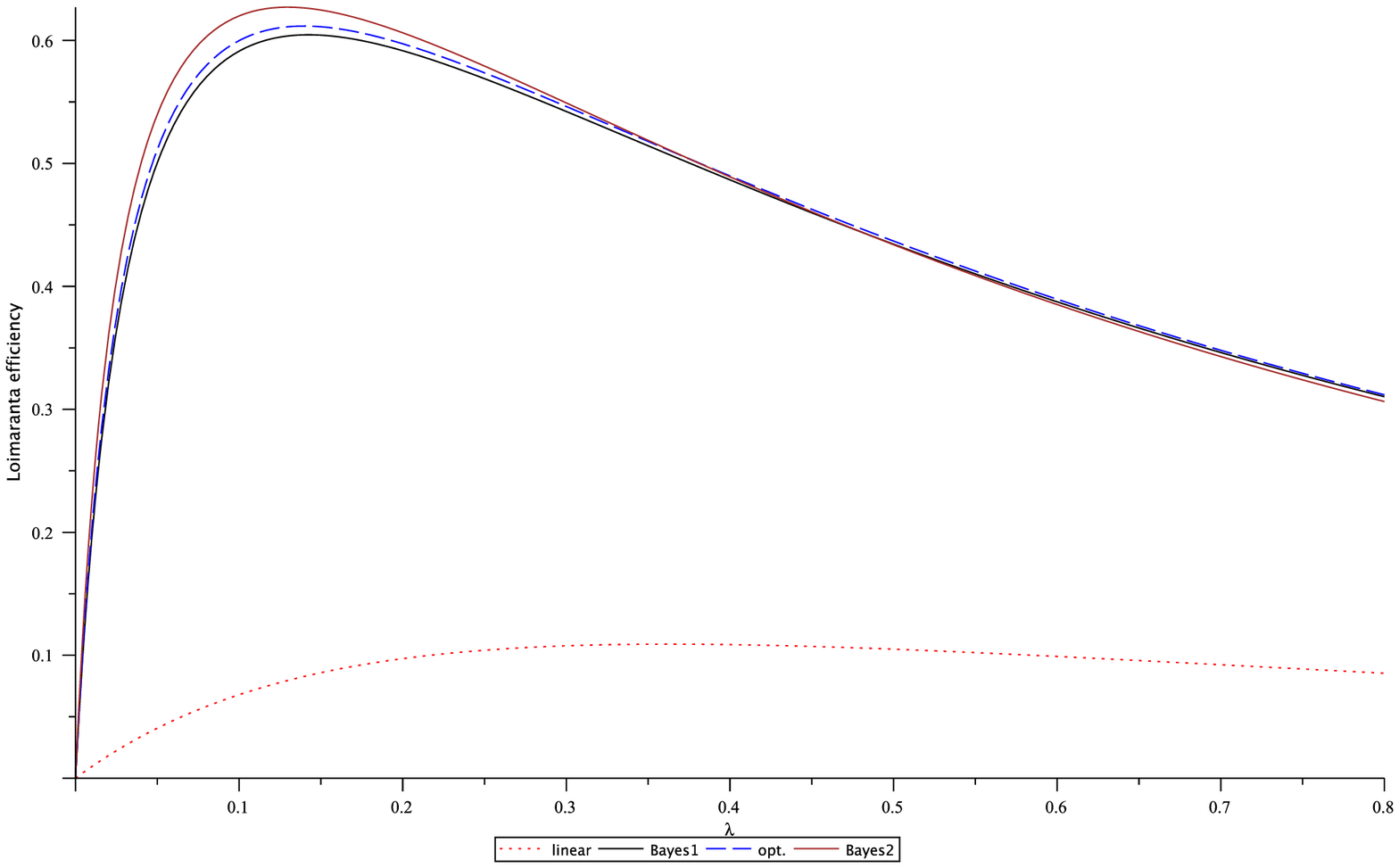}}\\ \subfigure[]{
\includegraphics[width=9cm,height=6cm]{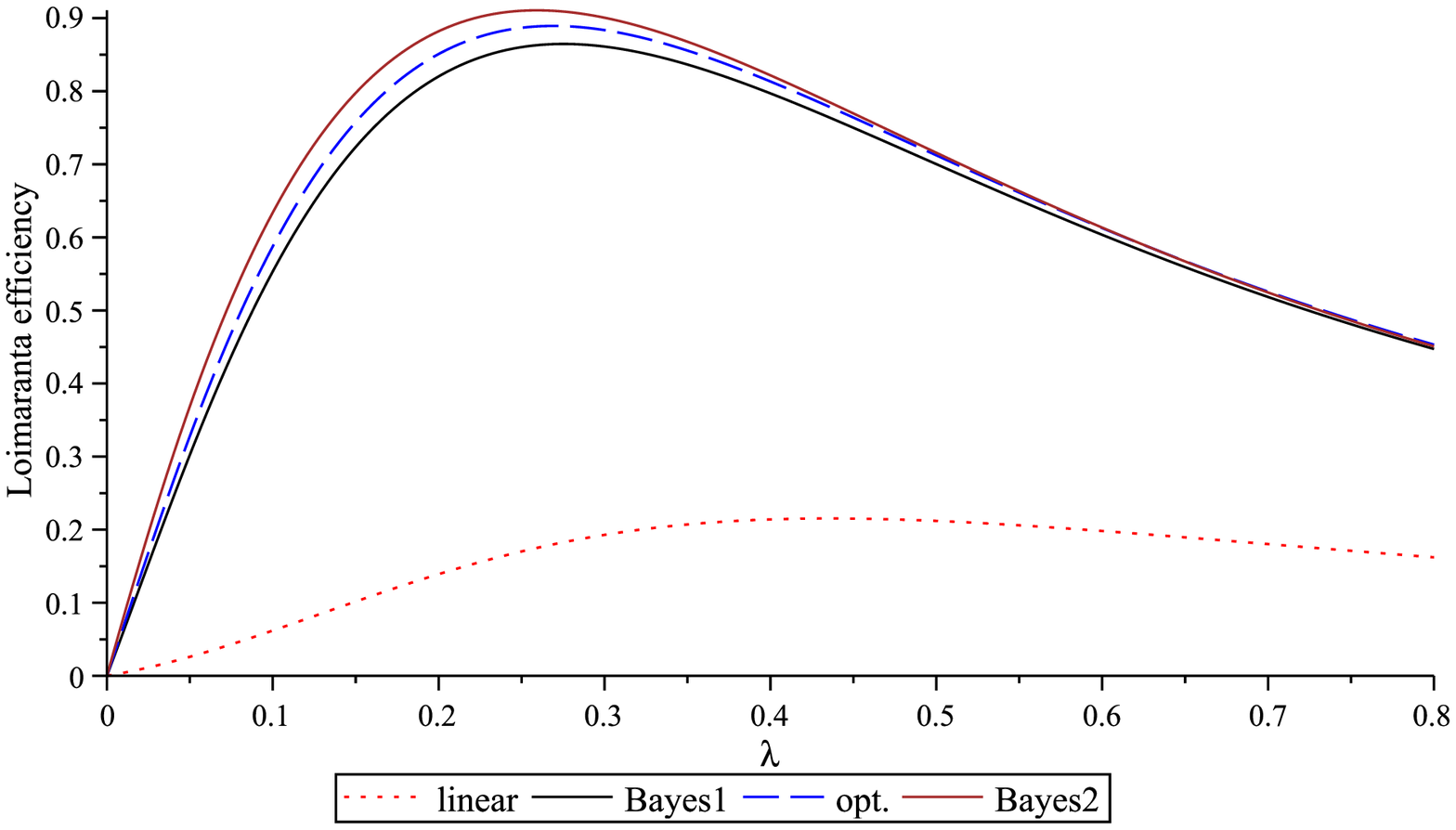}}\subfigure[]{
\includegraphics[width=9cm,height=6cm]{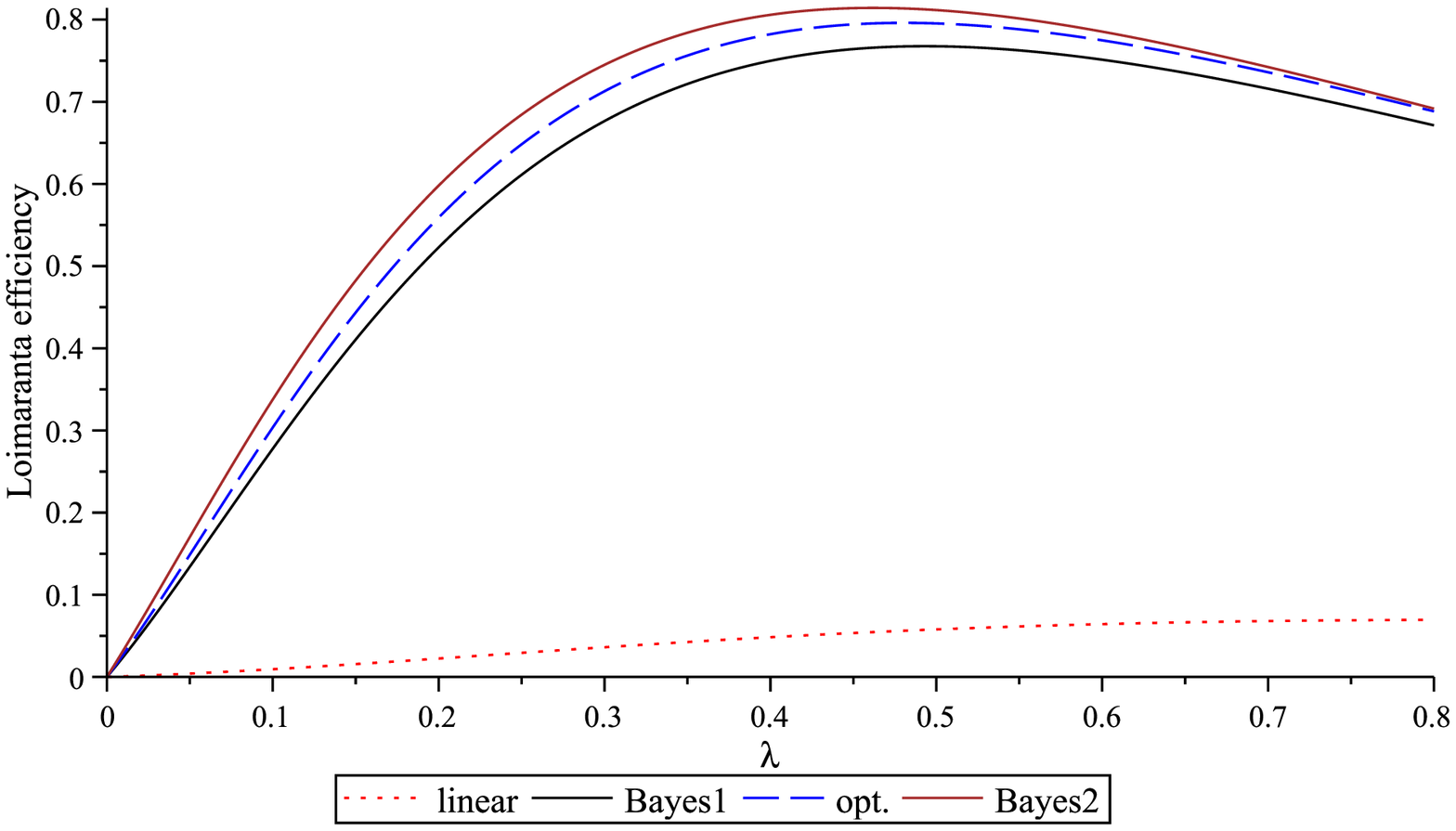}}\\ \subfigure[]{
\includegraphics[width=9cm,height=6cm]{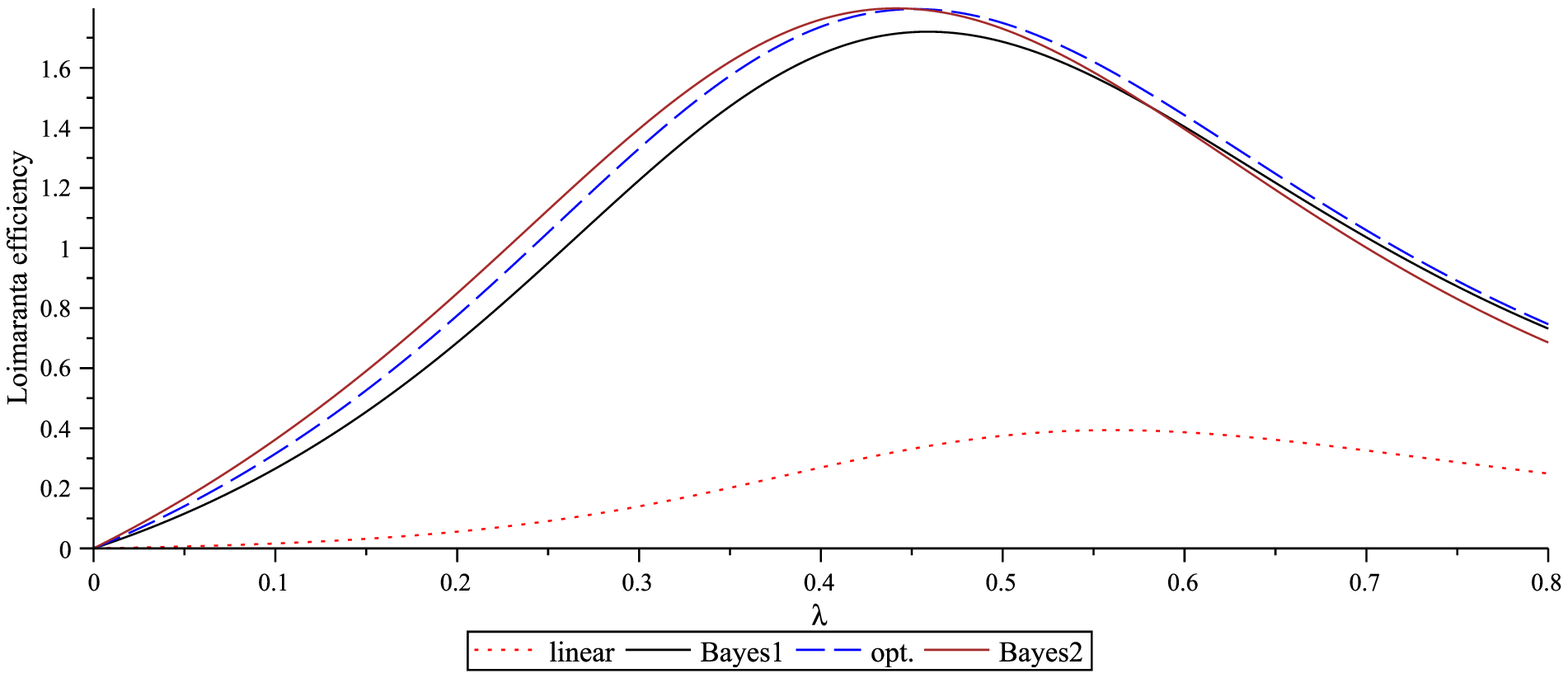}}\subfigure[]{
\includegraphics[width=9cm,height=6cm]{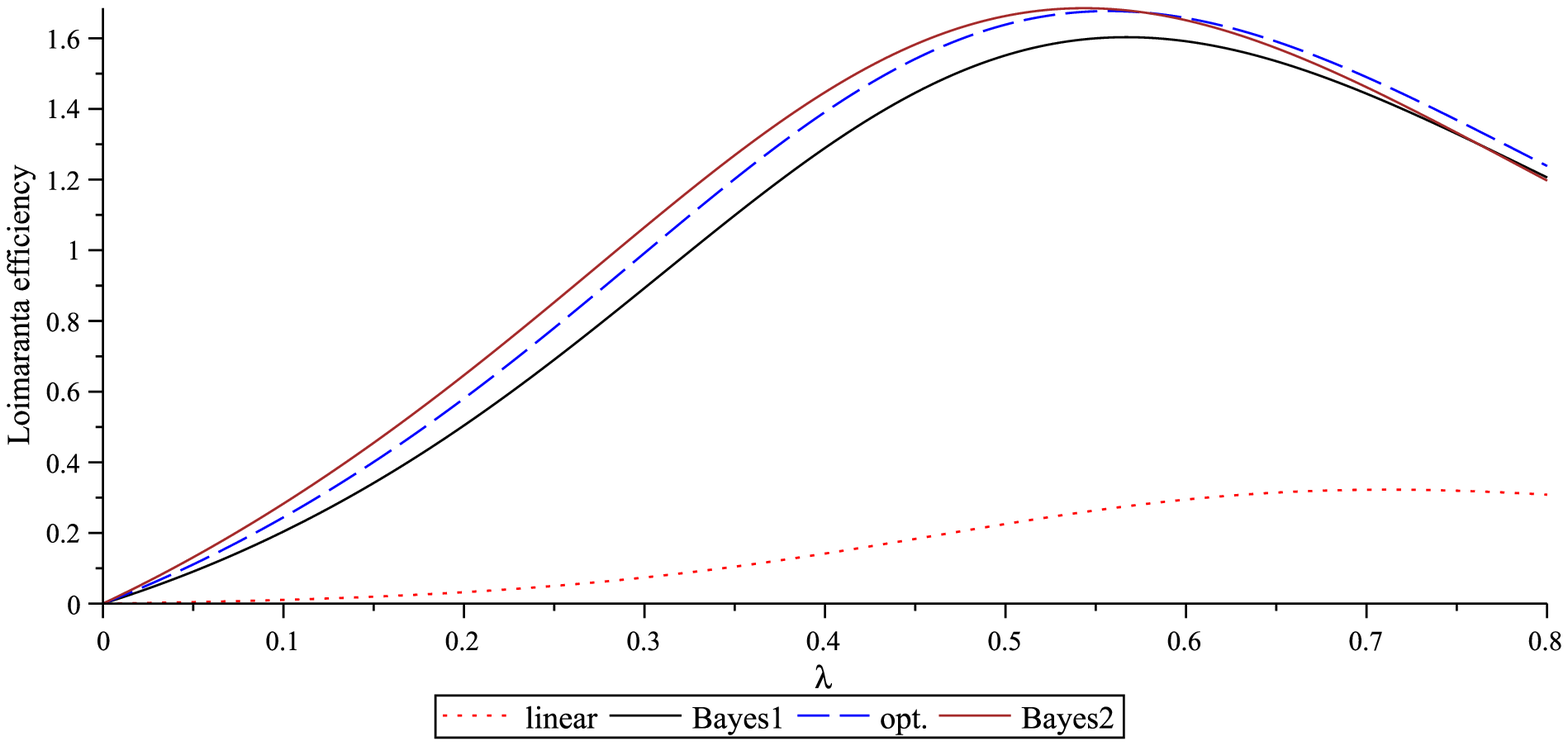}}
\caption{\scriptsize The Loimaranta efficiency for the four
relativity premiums under: Kenya's Bonus--Malus system, whenever $
N_l|\theta_l\sim Poisson(\theta_l)$ (Part a); Kenya's Bonus--Malus
system, whenever $ N_l|\theta_l\sim ZIPoisson(\theta_l)$ (Part b)
Hong Kong's Bonus--Malus system, whenever $ N_l|\theta_l\sim
Poisson(\theta_l)$ (Part c); Hong Kong's Bonus--Malus system,
whenever $ N_l|\theta_l\sim ZIPoisson(\theta_l)$ (Part d); and
Brazil's Bonus--Malus system, whenever $ N_l|\theta_l\sim
Poisson(\theta_l)$ (Part e); Brazil's Bonus--Malus system,
whenever $ N_l|\theta_l\sim ZIPoisson(\theta_l)$ (Part f) }
\end{figure}
\end{center}
Form Figure 1, one may observe that the Loimaranta efficiency of
the linear relativity premium is improved by using the
optimal linear relativity premium for all $\lambda\in[0,1].$
Moreover, for some $\lambda$, the Loimaranta efficiency of the
optimal linear relativity is relativity close to both Bayes
relativity premiums.
\subsection*{Base premium}
To derive the base premium, we suppose that random sample claim size
$X_1,\cdots,X_n,$ given risk parameter $\eta,$ has been
distributed according to one of the following four models. Moreover,
we suppose that risk parameter $\eta$ has prior distribution
$\pi_{1}(\cdot)$ {\it or} $\pi_{2}(\cdot)$ for the Bonus--Malus system
that has 7 or 6 levels, respectively.
\begin{description}
\item[Model 1:] Consider the mixture density function given by
Example \eqref{Example1_Bayes_mixture} with $n=20,$
$T_1=188.7745,$ $T_2=56.95046,$ $T_3=86422.7,$ and $T_4=691.2832.$
Moreover, suppose that the number of reported claims for a policyholder at
level $l,$ say $N_l,$ given risk parameter $\theta_l,$ is
distributed according to a Poisson distribution.
\item[Model 2:] Consider the mixture density function given by Example
\eqref{Example2_Bayes_mixture} with $n=200,$ $T_2=201.1964,$
$T_4=676.6038,$ and $x_{(1)}=0.3159083.$ Moreover, suppose that the number
of reported claims for a policyholder at level $l,$ say $N_l,$ given
risk parameter $\theta_l,$ is distributed according to a
Poisson distribution. \item[Model 3:] Consider the mixture density
function given by Example \eqref{Example1_Bayes_mixture} with
$n=20,$ $T_1=188.7745,$ $T_2=56.95046,$ $T_3=86422.7,$ and
$T_4=691.2832.$ Moreover, suppose that the number of reported claims for a
policyholder at level $l,$ say $N_l,$ given risk parameter
$\theta_l,$ is distributed according to a zero-inflated
Poisson distribution.
\item[Model 4:] Consider the mixture density function given by Example
\eqref{Example2_Bayes_mixture} with $n=200,$ $T_2=201.1964,$
$T_4=676.6038,$ and $x_{(1)}=0.3159083.$ Moreover, suppose that the number
of reported claims for a policyholder at level $l,$ say $N_l,$ given
risk parameter $\theta_l,$ is distributed according to a
zero-inflated Poisson distribution.
\end{description}
\begin{eqnarray*}
\pi_1(\eta) &=&
\frac{1}{7}Gamma(1,7)+\frac{1}{7}Gamma(3,7)+\frac{1}{7}Gamma(5,7)+\frac{1}{7}Gamma(7,7)+\frac{1}{7}Gamma(9,7)\\
&&+\frac{1}{7}Gamma(11,7)+\frac{1}{7}Gamma(13,7),\\
\pi_2(\eta) &=& \frac{1}{6}Gamma(1,6)+\frac{1}{6}Gamma(3,6)+\frac{1}{6}Gamma(5,6)\\
&&+\frac{1}{6}Gamma(7,6)+\frac{1}{6}Gamma(9,6)+\frac{1}{6}Gamma(11,6).
\end{eqnarray*}
Under the above conditions, Tables 5 to 7 report the base premium
along with the optimal relativity and level premiums.
\begin{landscape}\begin{scriptsize}\begin{center}Table 5: Relativity, Base and level premiums under Kenya's
Bonus--Malus system, for such four models.
\begin{tabular}{c c cc ccccccccc}
  \hline
  &  &Model 1&  &&Model 2&  &&Model 3&  &&Model 4&\\
  \cline{2-13}
  l & $r_l$ & Base & $Premium_l$ & $r_l$ & Base & $Premium_l$ & $r_l$ & Base & $Premium_l$ & $r_l$ & Base & $Premium_l$ \\
  \hline
  1 & 0.139 & 2.705 & 0.376 & 0.139 &  0.984 & 0.137 & 0.142 & 2.705 & 0.384 & 0.142 &  0.984 & 0.140 \\
  2 & 0.436 & 2.705 & 1.179 & 0.436 &  0.984 & 0.429 & 0.440 & 2.705 & 1.190 & 0.440 &  0.984 & 0.433 \\
  3 & 0.732 & 2.705 & 1.983 & 0.732 &  0.984 & 0.721 & 0.737 & 2.705 & 1.996 & 0.737 &  0.984 & 0.726 \\
  4 & 1.029 & 2.705 & 2.786 & 1.029 &  0.984 & 1.014 & 1.035 & 2.705 & 2.802 & 1.035 &  0.984 & 1.019 \\
  5 & 1.326 & 2.705 & 3.590 & 1.326 &  0.984 & 1.306 & 1.333 & 2.705 & 3.608 & 1.333 &  0.984 & 1.313 \\
  6 & 1.622 & 2.705 & 4.393 & 1.622 &  0.984 & 1.598 & 1.631 & 2.705 & 4.415 & 1.631 &  0.984 & 1.606 \\
  7 & 1.919 & 2.705 & 5.196 & 1.919 &  0.984 & 1.890 & 1.928 & 2.705 & 5.221 & 1.928 &  0.984 & 1.899 \\
  \hline
\end{tabular}\\
Table 6: Relativity, Base and level premiums under Hong Kong's
Bonus--Malus system, for such four models.
\begin{tabular}{c c cc ccccccccc}
  \hline
  &  &Model 1&  &&Model 2&  &&Model 3&  &&Model 4&\\
  \cline{2-13}
  l & $r_l$ & Base & $Premium_l$ & $r_l$ & Base & $Premium_l$ & $r_l$ & Base & $Premium_l$ & $r_l$ & Base & $Premium_l$ \\
  \hline
  1 & 0.917 & 2.719 & 2.480 & 0.917 & 0.983 & 0.902 & 0.985 & 2.719 & 2.664 & 0.985 & 0.983 & 0.969 \\
  2 & 1.036 & 2.719 & 2.802 & 1.036 & 0.983 & 1.019 & 1.041 & 2.719 & 2.813 & 1.041 & 0.983 & 1.023 \\
  3 & 1.154 & 2.719 & 3.124 & 1.154 & 0.983 & 1.137 & 1.096 & 2.719 & 2.962 & 1.096 & 0.983 & 1.077 \\
  4 & 1.273 & 2.719 & 3.446 & 1.273 & 0.983 & 1.254 & 1.152 & 2.719 & 3.111 & 1.152 & 0.983 & 1.132 \\
  5 & 1.392 & 2.719 & 3.768 & 1.392 & 0.983 & 1.371 & 1.208 & 2.719 & 3.260 & 1.208 & 0.983 & 1.186 \\
  6 & 1.510 & 2.719 & 4.090 & 1.510 & 0.983 & 1.488 & 1.263 & 2.719 & 3.408 & 1.263 & 0.983 & 1.240 \\
  \hline
\end{tabular}\\
Table 7: Relativity, Base and level premiums under Brazil's
Bonus--Malus system, for such four models.
\begin{tabular}{c c cc ccccccccc}
  \hline
  &  &Model 1&  &&Model 2&  &&Model 3&  &&Model 4&\\
  \cline{2-13}
  l & $r_l$ & Base & $Premium_l$ & $r_l$ & Base & $Premium_l$ & $r_l$ & Base & $Premium_l$ & $r_l$ & Base & $Premium_l$ \\
  \hline
  1 & 0.141 & 2.705 & 0.381 & 0.141 & 0.984 & 0.139 & 0.142 & 2.705 & 0.384 & 0.142 & 0.984 & 0.140 \\
  2 & 0.499 & 2.705 & 1.350 & 0.499 & 0.984 & 0.491 & 0.495 & 2.705 & 1.339 & 0.495 & 0.984 & 0.487 \\
  3 & 0.846 & 2.705 & 2.318 & 0.846 & 0.984 & 0.843 & 0.848 & 2.705 & 2.294 &  0.848 & 0.984 & 0.834 \\
  4 & 1.199 & 2.705 & 3.287 & 1.199 & 0.984 & 1.196 & 1.201 & 2.705 & 3.249 & 1.201 & 0.984 & 1.182 \\
  5 & 1.551 & 2.705 & 4.255 & 1.551 & 0.984 & 1.548 & 1.555 & 2.705 & 4.204 & 1.555 & 0.984 & 1.529 \\
  6 & 1.904 & 2.705 & 5.223 & 1.904 & 0.984 & 1.900 & 1.908 & 2.705 & 5.158 & 1.908 & 0.984 & 1.876 \\
  7 & 2.257 & 2.705 & 6.192 & 2.257 & 0.984 & 2.252 & 2.261 & 2.705 & 6.113 & 2.261 & 0.984 & 2.224
 \\
  \hline
\end{tabular}
\end{center}\end{scriptsize}\end{landscape}
\section{Conclusion and suggestions}
This article designs an optimal Bonus--Malus System by evaluating
relativity and base premiums. To estimate the relativity premium,
this article considers a class of linear relativity premiums and
determines an optimal premium within this class such that the
estimator is simultaneously close to both possible Bayes
relativity premiums. The base premium is evaluated under a
Bayesian framework and two finite mixture models for both random
claim size and risk parameter $\eta.$ The Loimaranta efficiency
shows that the efficiency of the new linear relativity premium is
drastically improved compared with the ordinary relativity
premium.

\section*{Acknowledgments}
The authors thank professor Jan Dhaene for his useful comments and
suggestions on an earlier version of this manuscript.

\end{document}